# RELARM: A rating model based on relative PCA attributes and k-means clustering

Elnura Irmatova [1] [2]

## Abstract

Following widely used in visual recognition concept of relative attributes, the article establishes definition of the relative PCA attributes for a class of objects defined by vectors of their parameters. A new rating model (RELARM) is built using relative PCA attribute ranking functions for rating object description and k-means clustering algorithm. Rating assignment of each rating object to a rating category is derived as a result of cluster centers projection on the specially selected rating vector. Empirical study has shown a high level of approximation to the existing S & P, Moody's and Fitch ratings.

**Key words:** rating model, relative PCA attribute, credit rating, principal component analysis, k-means clustering.

[1] Russian Presidential Academy of National Economy and Public Administration (RANEPA), Faculty of Finance and Banking, Interdepartmental laboratory of financial and economic research. Prospect Vernadskogo, 82, Moscow, Russian Federation 119571. E-mail: elnura-irmatova@mail.ru

[2] This research was conducted within Interdepartmental laboratory of financial and economic research (RANEPA, Faculty of Finance and Banking)

# I. Introduction

Credit rating agencies play an important role in providing financial markets with indicative and prognostic information leading to increase of market efficiency. One of the key factors when choosing the rating agency is the creditor's confidence in its estimates, so that each agency is struggling to improve their assessments transparency for the rating end-users. It should be noted that the mentioned problem is often referred to the newly created agencies while they are building the agency reputation. Companies, investors and stakeholders thoroughly examine the evaluation principles of their rating methodologies at that time. In this regard, when establishing the rating methodology you might face a problem of the result objectivity. In this context, a rating methodology objectivity means a minimum presence of model factors evaluated solely by expert judgment.

It is important to mention that one of the rating agency's priority is to choose the type of model that describes a rating object with a minimum use of subjective expert factors.

Usually credit rating models are scoring type models. A scoring model involves obtaining the integral numerical index based on quantitative and qualitative parameters where each of them has a certain influence on the creditworthiness of a rating object expressed by a specific weighting factor. Such models can be built using econometric tools (ranging from simple linear regression to logit and probit models), multiplicative discriminant analysis, neural networks, support vector machine techniques or on the basis of expert judgment [1], [2], [3], [4].

The difference between the rating model types is in various approaches to rating object data base processing, diversity in tools for model establishment and variation in obtaining of weight coefficients.

For example, factor weights in a model based on expert judgment are determined by an expert community, however the other model types often involve special model training to find the right weights. It should be noted that model factors are always an expert community matter of choice and their selection is carried out in accordance with a concept of inclusion the majority of relevant to analyzed rating object characteristics.

*<u>Problem formulation</u>*

Credit rating agencies commonly use evaluation models with a high degree of expert component and some of them incorporate models based on econometric tools. Here the expert part is to identify the degree of factors' influence on the final rating as well as to determine intervals of model parameters for the following point assignment.

The main difficulty of such rating methodologies is that expert methods are not always transparent to the rating end users as there might exist plenty opinions about a specific weight of an indicator or a model part as well as options for their numerical values interpretation. In addition, a model construction and calibration based solely on expert judgment practically can lead to bias in the final rating. For example, the expert selection of factor weights may cause significant inaccuracy in resulting rating assignment due to a possible high level of interdependence between the analyzed indicators.

On the other hand, the choice of an econometric method for model construction or simulation of a neural network system requires a broad rating history database for model training and adjustment, otherwise, there might occur substantial errors in rating prediction and assignment. Thus, the lack of access to a full rating history database is the second major difficulty in building an accurate automated rating model.

Furthermore, it should be noted, that existing in practice and described in scientific literature rating models, in fact, are built regardless of a rating object's interdependencies, although practically such kind of difficulties can be coped with an expert rating committee decisions.

*<u>A new rating model based on relative PCA attribute ranking functions</u>*

In this paper, we introduce a new rating model based on relative PCA (principal component analysis) attribute ranking functions and k-means clustering (***RELARM: Relative Attributes Rating Model)*** with the following distinctive features:

1. Rating assignment taking into account comprehensive rating object interdependencies;

2. Simplicity of model training and calculation on small but relevant data array.

The proposed model aims to determine a rating object creditworthiness (financial strength/ stability) based on the principle of "living organism" where each element change (even very small) causes certain reflection on the state of other analyzed system objects. Incorporation of a new for rating models definition of the *relative PCA attributes* provides the most comprehensive description of analyzed rating object characteristics. It should be emphasized that obtained by RELARM rating results are robust if model parameters are properly chosen.

The described above the second model's feature signifies a possibility to train RELARM on the 1-2 years relevant to rating object data, so that it is becoming unnecessary to use large training samples (e.g. 10, 20, 30 years etc.). It should be noted that model adjustment is going to be more accurate with an increase of data horizon. However, adequate model results might be obtained with a minimum years of relevant data.

Therefore, proposed in the article new rating model (RELARM) is aimed to provide credit rating agencies with a rating methodology based on the principles of rating objectivity and transparency even when they experience limitations on training data sample.

In Section II, we describe related to the paper works. In Section III, we present the theoretical concept of RELARM. In Section IV we show an empirical study and section V concludes.

**II. Related works**

Nowadays the credit rating topic is very popular in scientific literature. There exist a large amount of works on such topics as rating modeling, credit scoring and determination of credit quality.

The current section contains a small part of the research works on actual rating modeling however these papers clearly reflect the relevance of the chosen theme as well as they show a variety of tools possible to apply.

We conditionally divide the existing papers on rating modeling into 3 types (by the type of tools they suggest to apply):

— models using expert judgment,

- models using econometric tools,
- models using machine learning techniques.

*Models using expert judgment*

In practice, credit rating agencies often use models based on expert judgment.

One of the most frequent methods used to determine influence of rating model factors with the use of expert opinion is called Analytic Hierarchy Process (AHP). In [5] AHP is incorporated in conjunction with the grey hierarchy evaluation model. Another approach is presented in [6] where the credit quality is explored with the improved grey relation analysis (GRA), AHP and TOPSIS (the Technique for Order Performance by Similarity to Ideal Solution). Paper [7] proposes credit default risk evaluation based on Dynamic Multiple Criteria Decision Making Model, AHP and UTADIS (UTilites Additives DIScriminantes) method for the final scores computation and ranking.

*Models using econometric tools*

Regression models are the second most popular models type applied by rating agencies in practical field. It should be noted that the most commonly used form of regression in credit ratings is a logistic regression. For example, the general principles can be found in [8] where the model is built and tested with inclusion of quality parameters. In [9] a credit scoring model based on fuzzy logistic regression is constructed. Another variant is presented in [10] where authors used an ordinal regression approach to construct a rating model for sovereign creditworthiness assessment.

*Models using machine-learning techniques*

In the last decade machine-learning techniques such as neural networks and support vector machines were intensively developed. They have been widely used in image recognition systems as well as in theoretical credit rating modeling, determination of credit risk and quality. Papers [11], [12], [13] propose methods using support vector machine (SVM) technique and its modifications. A neural network rating model is presented in [14].

One can also find implementation of principal component analysis (PCA) in credit rating modeling [15], [16], [17], [18].

**III. RELARM theoretical description**

Current section reveals the theoretical concept of a new rating model based on relative PCA attribute ranking functions and k-means clustering. Here is presented in detail a step by step description of RELARM construction.

RELARM is based on 4 phases:

I. Normalization of input data – unification of initial model parameters for their comparison using linear scaling method.

II. Relative PCA attribute ranking functions calculation, normalized parameter vectors mapping in the space of relative PCA attribute ranking function values and formation of the rating vector.

III. k-means clustering of the relative PCA attribute ranking function space vector values with cluster centers obtaining.

IV. Rating assignment of analyzed rating objects by projection on the rating vector.

In 3.1 we describe normalization of factors procedure. Then in 3.2 we give a definition of relative PCA attributes and their ranking functions for rating object description. The mapping in the space of relative PCA attribute ranking functions values is defined. In 3.3 we show k-means clustering application to the proposed model.

*3.1 Normalization of RELARM input parameters*

Suppose that a rating model consists of N factors and M rating objects. We apply a linear scaling method in order to standardize rating model parameters for their comparability.

Let $p_{ij}$, $i \in [M], j \in [N]$ denote the initial value of the *j-th* parameter of the *i-th* rating object. We define a normalized value $b_{ij}$ of $p_{ij}$, where $i \in [M], j \in [N]$, depending on the *j-th* factor's influence on the model property studied.

If an increase of $p_{ij}$ index value has a positive impact on the final analyzed property, the formula becomes:

$$b_{ij} = \frac{p_{ij} - \min_i p_{ij}}{\max_i p_{ij} - \min_i p_{ij}}, i \in [M], j \in [N]. \qquad (3.1)$$

If a model parameter increase has a negative effect on the final rating, then normalized value $b_{ij}$ is calculated as:

$$b_{ij} = \frac{\max_i p_{ij} - p_{ij}}{\max_i p_{ij} - \min_i p_{ij}}, i \in [M], j \in [N]. \qquad (3.2)$$

As a result of (3.1) and (3.2) each rating object is described by a (1 x n) dimension row vector of normalized parameters:

$$b_i^T = (b_{i1}, \ldots, b_{iN}) \in [0,1]^N, i \in [M]. \qquad (3.3)$$

Let

$$B \coloneqq \{b_i\}, i \in [M]. \qquad (3.4)$$

denote a set of normalized parameters.

### 3.2 Rating object characteristic using relative PCA attribute ranking function values

Current paragraph reveals the second phase of RELARM construction, namely the mapping of normalized vectors $b_i$, $i \in [M]$ to the space of relative PCA attribute ranking function values. The concept of attributes is widely used in image recognition algorithms. It is most often presented in recognition using binary properties, which predicts a presence or an absence of a specific attribute (e.g. smiles on photos, determination of a landscape type etc.). However, the use of such algorithms has certain restrictions and often leads to ambiguous recognition or total disregard of a characteristic. Later in paper [19] it is proposed an application of *relative attributes* providing semantically more rich method for object description, which uses objects features comparison in relation to each other. The concept of relative attributes provides a relative strength of specified features presence of an object compared to other objects.

*Definition of a relative PCA attribute and its ranking function*

Let

$$w_k = \begin{pmatrix} w_{1k} \\ \vdots \\ w_{Nk} \end{pmatrix} \qquad (3.5)$$

denote $l_1$-normalized PCA components of the set $B$ (3.4) with principal component variances $\lambda_k, \lambda_1 \geq \lambda_2 \geq \cdots \geq \lambda_N$.

**Definition.** Let *p-th* relative PCA attribute of vector $b_i \in B$ $i = 1, 2, \ldots M$ be a vector $A_{ip}$:

$$A_{ip} = (b_{i1}w_{1p}, b_{i2}w_{2p}, \ldots, b_{iN}w_{Np}), \quad p = 1, \ldots, N. \qquad (3.6)$$

Further we can also name the *p-th* relative PCA attribute of a vector as the **_p-th_ main attribute vector.** In accordance with the concept presented in [19] we say that the *p-th* main attribute has a stronger presence in vector $b_i$ than in vector $b_j$, if $l_1$-norm of vector $A_{ip}$ is greater than $l_1$-norm of vector $A_{jp}$:

$$\sum_{k=1}^{N} b_{ik}|w_{kp}| \geq \sum_{k=1}^{N} b_{jk}|w_{kp}| \qquad (3.7)$$

Therefore, the ranking vector for *p-th* main attribute is the vector $\tilde{w}_p^T$:

$$\tilde{w}_p^T = (|w_{1p}|, \ldots, |w_{Np}|), \qquad (3.8)$$

and the ranking function is defined by formula:

$$r_p(b_i) = \tilde{w}_p^T b_i \qquad (3.9)$$

([19]).

We define $N \times d$ matrix $W$ as:

$$W = \begin{pmatrix} |w_{11}| & \cdots & |w_{1d}| \\ \vdots & \ddots & \vdots \\ |w_{N1}| & \cdots & |w_{Nd}| \end{pmatrix}, \qquad (3.10)$$

where the number of principal components $d$ is determined to avoid the influence of «data noise». We recommend to take the number of principal components $d$ providing approximately 95% of data information.

We define the rating vector Λ as:

$$\Lambda := (\lambda_1, \lambda_2, \ldots, \lambda_d). \tag{3.11}$$

Let $f: B \to \mathbb{R}^d$ be a map of the set B to the space $\mathbb{R}^d$ of relative attribute ranking functions values defined by formula:

$$a_i^T := f(b_i^T) = b_i^T \times W. \tag{3.12}$$

Here:

$$a_i^T = (r_1(b_i),\ r_2(b_i), \ldots,\ r_d(b_i)) = \left(\sum_{k=1}^{N} b_{ik}|w_{k1}|, \sum_{k=1}^{N} b_{ik}|w_{k2}|, \ldots, \sum_{k=1}^{N} b_{ik}|w_{kd}|\right) \tag{3.13}$$

For the *i-th* rating object each component of vector $a_i^T$ indicates the degree of influence of object's parameter changes with respect to the corresponding principal component.

### *3.3 Rating assignment using k-means clustering*

In this paper, k-means algorithm [20], [21] is used for the final rating objects classification to specific rating categories.

It should be noted that we take Euclidean distance as a distance measure between data points but practically it is possible to use other existing distance options suitable for a particular issue.

Partitioning of rating objects to rating classes includes 4 stages:

1. K-means clustering algorithm application to obtained in 3.2 vectors of relative PCA attribute ranking function values (3.13) . The output cluster centers we denote by $CC_q, q \in (1,2,\ldots,k)$;

2. Cluster centers projection on the rating vector Λ (3.11);

3. Ranking of centers projection on the rating vector in descending order (the higher the value the better credit quality);

4. Rating assignment on the basis of 1 and 3.

Module of projection of the *q-th* cluster center on the rating vector Λ (3.11) is calculated as follows:

$$PR_q = |(CC_q, \Lambda)|, q \in (1,2,\ldots,k). \tag{3.14}$$

**Note.** We can assume intuitively that the *l-th* value of the rating vector $\Lambda$ is proportional to probability that comparison by the *l-th* main attribute ranking function is correct. Thus, $PR_q$ can be considered as analogue of module of a rating object's expected value.

Finally, we rank obtained values $PR_q$, $q \in (1,2,...,k)$ in descending order and form a rating list containing analyzed rating objects with respect to assigned rating classes.

**IV. RELARM empirical study**

In current section, we present an empirical example of RELARM application based on the theoretical concept proposed in the previous section.

For the purpose of our experiment, it was decided to build a test sovereign credit rating assessment model using RELARM approach. The main reasons for the choice of countries data for model construction were:

— firstly, it is possible to collect a large number of county relevant indicators in public databases;

— secondly, sovereign credit rating can be obtained with a greater emphasis on the economic and financial indicators and a smaller part containing expert based variables. However, for example, an enterprise or financial institution credit rating always comprises a significant number of qualitative indicators.

The second part is especially important for an experiment model because in this case minimization of expert opinion is necessary for a greater result neutralization, which makes possible to compare the final test model rating with the real sovereign credit ratings.

Finally, section concludes the adequacy of the results obtained in comparison with the assigned Standard & Poor's[3], Moody's[4] и Fitch[5] rating agencies sovereign credit ratings.

*4.1 Data and model parameters*

---

[3] https://www.standardandpoors.com/
[4] https://www.moodys.com/
[5] https://www.fitchratings.com/

The initial and fundamental phase of a rating model construction is the selection of the most relevant parameters characterizing a rating object as full as possible within evaluation of a certain property (e.g. creditworthiness, reliability, efficiency, quality, etc.). Criteria choice largely influences the final model result providing minimum outliers if they were picked correctly.

Consequently, for realization of an experiment we selected a part of indicators for sovereign credit rating assessment used by Standard & Poor's, Moody's и Fitch rating agencies. It is also important to mention that our criteria choice was based on the possibility to collect the latest data.[6]

Thus, 9 financial and economic parameters, fundamental for economy description, were selected. The description is presented in Table 1.

In addition to financial and economic indicators there also included an expert factor with a distribution [0,1] (likewise the other normalized parameters). It is supposed to reflect an expert opinion on the country's economy strength, a possibility to gain financial support from other countries, «soft power» as well as prediction of economy's power. In order to conduct an experiment 30 countries were selected.

Data for the selected criteria were taken from public statistical databases: the World Bank, data from country's statistical agencies, central banks and other sources.

*4.2 Calculations*

Once the evaluation criteria were selected, a country list was compiled and all the relevant data was collected, normalization of the input data should be made. However, before that step to be done, it is necessary to determine the influence of each factor on the final property studied, namely creditworthiness (Table 2).

Next, normalized values for each factor are calculated according to formulae (3.1) and (3.2) for factors with positive and negative influence respectively. Table 3 shows an example of normalized values calculation. Matrix of normalized countries' indicators values is presented in Table 5. As it was mentioned above, we also inserted a specific expert factor (with [0,1] distribution) in addition to selected economic and financial indicators.

---

[6] We took mostly 2016 year data but if there was no current date on a factor in open access and we suggest the indicator is important for the model, we included data from previous years.

Based on the matrix of normalized country indicators values (Table 4) with an added expert factor, the number of principal components $d$ and matrix $W$ (3.10) are determined. The calculated matrix $W$ is presented in Table 5.

Next, the mapping in the space $R^d$ of relative attribute ranking functions values (3.12) is performed. It should be noted, that in this particular case we selected 6 principal components which ensure preservation of 96% data information. Also the rating vector $\Lambda$ (3.11) was obtained (Table 6).

Therefore, we have 30 vectors in the $R^6$ space corresponding to each country of a sample and then the algorithm described in 3.3 is performed. As we consider a small sample for the experiment, it was decided to take $k$=7 for k-means clustering which means that we exclude rating subcategories and consider the following classes: AAA, AA, A, BBB, BB, B, CCC. It should be mentioned that Euclidean distance was selected for the clustering algorithm. K-clustering allowed to make homogenous groupings as well as to compute cluster centers presented in Table 7.

Next, according to algorithm from 3.3, modules of projections $PR_q$ were found, then ranking of obtained cluster projections was made and, finally, rating assignment to each country was performed.

The resulting RELARM rating was compared with existing country ratings on 31.07.2016 (source: Thomson Reuters Eikon) assigned by S & P, Moody's and Fitch rating agencies. We consider a country's RELARM rating to be a success result, if it matches **_one_** of the real ratings of the above-mentioned rating agencies. Since we divided the test sample into 7 rating categories, excluding subcategories, the test model country's result is identical to an existing country rating if it complies with the general rating category (Ex .: the country's rating of the test model is AA and this country has Fitch AA + : RELARM rating matches the real one).

Thus, an experiment model based on RELARM method showed results presented in Table 8. It can be seen from the Table 8 that RELAM method provided 86% approximation to real ratings. Moreover, such an effective result was gained using only 10 model factors. In addition, thorough indicators selection in conjunction with a correct model and k-means clustering adjusting can provide an opportunity to refine the results. It should be noted, the resulting country distribution to rating categories is a ***recommendation*** to a rating agency's rating committee for a rating assignment, however the final decision on the rating level is always made by experts.

The result obtained using RELARM approach is adequate and appropriate for implementation in rating agencies practical activities.

**IV. Conclusion**

A new rating model based on the relative PCA attributes ranking functions and k-means clustering was proposed.

RELARM provides the most comprehensive rating objects description by introducing the relative PCA attributes concept for object's parameters vector.

An empirical study showed high approximation level to actual countries credit ratings, which proved the RELARM approach to be adequate. A wise model factors selection and correct clustering mechanism adjusting can ensure results accuracy. Simplicity of model training and results computation at the same time are the basis for RELARM implementation in the practical field of rating agencies.

**Acknowledgement**

**Table 1. Description of experiment indicators**

| Indicator | Short description | Rating agency that uses this indicator[7] |
|---|---|---|
| **1. Average GDP growth** (for 7 years: 6 previous periods and the current data) | Generally, GDP growth indicates an economic development, whereas a decrease leads to economic recession. | Moody's Fitch |
| **2. Global Competitiveness Index (World Economic Forum)** | Economic competitiveness, which is the country's ability to provide high level of well-being for citizens according to the World Economic Forum. | Moody's |
| **3. GDP per capita** | In general terms, it shows the level of economic development and prosperity. | Moody's Fitch |
| **4. Government debt/GDP** | In general terms, the higher level of public debt leads to the greater country risk of default. Nevertheless, such relationship is not linear, and it should be considered in conjunction with other factors. | Moody's Fitch |
| **5. Budget surplus (+) or deficit (-) (% of GDP)** | Generally, a large budget deficit shows the fiscal policy inconsistency and, if other things being equal, it may lead to an increase in the debt burden. | Fitch |
| **6. Inflation level** | In the general sense, the low inflation level and economic growth lead to a more stable credit position and vice versa. | Moody's Fitch |
| **7. Inflation volatility** (for 5 years: 4 previous periods and the current data) | Generally, high inflation volatility suggests monetary policy inconsistency and vice versa. | Moody's |
| **8. (Current account balance + Foreign direct investments)/GDP** | Indicator characterizes the ability to cover a current account deficit - the higher index value, the lower probability of a debt burden increase and, thus, the lower chance of creditworthiness reduction. | Moody's Fitch |
| **9. Reserves** | Country's «buffer» in various financial or economic shocks. Generally, higher values indicate a higher country's stability in such crises. | Fitch |

---

[7] According to S&P, Moody's and Fitch sovereign credit rating methodologies, July 2016

**Table 2. Indicator influence on countries' creditworthiness.**

| Indicator | Influence |
|---|---|
| 1. Average GDP growth | Positive |
| 2. Global Competitiveness Index (World Economic Forum) | Positive |
| 3. GDP per capita | Positive |
| 4. Government debt/GDP | Negative |
| 5. Budget surplus (+) or deficit (-) (% of GDP) | Positive |
| 6. Inflation level | Negative |
| 7. Inflation volatility | Negative |
| 8. (Current account balance + Foreign direct investments)/GDP | Positive |
| 9. Reserves | Positive |

**Table 3. Example of normalized values calculation**

| Country: Russia | | | | |
|---|---|---|---|---|
| **Indicator** | Input value | The minimum value for the sample of 30 countries | The maximum value for the sample of 30 countries | Normalized value |
| Global Competitiveness Index (World Economic Forum) | 4.44 | 3.3 | 5.76 | 0.4634 |
| Inflation level | 7.5% | -1.3 | 180.9 | 0.9517 |

**Table 4. Matrix of normalized indicator values for the sample of 30 countries**

| Country | GDP growth | WEF Competitiveness | GDP per Capita | Gov debt/GDP | Budget balance/GDP | Inflation level | Inflation Volatility | (CAB+FDI)/GDP | Reserves |
|---|---|---|---|---|---|---|---|---|---|
| Switzerland | 0.45 | 1.00 | 1.00 | 0.86 | 0.67 | 1.00 | 1.00 | 1.00 | 0.17 |
| Norway | 0.43 | 0.86 | 0.93 | 0.87 | 1.00 | 0.97 | 0.99 | 0.57 | 0.02 |
| United States | 0.49 | 0.94 | 0.68 | 0.55 | 0.52 | 0.99 | 1.00 | 0.20 | 0.03 |
| Germany | 0.47 | 0.91 | 0.49 | 0.69 | 0.71 | 0.99 | 0.99 | 0.48 | 0.02 |
| Austria | 0.42 | 0.74 | 0.52 | 0.63 | 0.60 | 0.99 | 0.99 | 0.31 | 0.00 |
| Finland | 0.37 | 0.87 | 0.50 | 0.73 | 0.51 | 0.99 | 0.99 | 0.40 | 0.00 |
| United Kingdom | 0.49 | 0.87 | 0.52 | 0.62 | 0.41 | 0.99 | 0.99 | 0.08 | 0.04 |
| France | 0.41 | 0.74 | 0.43 | 0.58 | 0.46 | 0.99 | 1.00 | 0.22 | 0.02 |
| Belgium | 0.42 | 0.77 | 0.48 | 0.54 | 0.52 | 0.98 | 0.99 | 0.05 | 0.00 |
| Korea | 0.61 | 0.69 | 0.31 | 0.85 | 0.49 | 0.99 | 1.00 | 0.39 | 0.11 |
| China | 1.00 | 0.65 | 0.06 | 0.81 | 0.53 | 0.98 | 1.00 | 0.34 | 1.00 |
| Czech Republic | 0.46 | 0.57 | 0.18 | 0.83 | 0.65 | 0.99 | 0.99 | 0.24 | 0.02 |
| Japan | 0.41 | 0.88 | 0.38 | 0.00 | 0.32 | 1.00 | 0.99 | 0.22 | 0.36 |
| Estonia | 0.59 | 0.59 | 0.18 | 0.96 | 0.69 | 1.00 | 0.97 | 0.21 | 0.00 |
| Saudi Arabia | 0.70 | 0.72 | 0.22 | 1.00 | 0.53 | 0.97 | 1.00 | 0.57 | 0.18 |
| Mexico | 0.58 | 0.40 | 0.07 | 0.82 | 0.47 | 0.98 | 1.00 | 0.21 | 0.05 |
| Kazakhstan | 0.68 | 0.48 | 0.09 | 0.90 | 0.60 | 0.90 | 0.92 | 0.34 | 0.01 |
| Bulgaria | 0.44 | 0.41 | 0.04 | 0.88 | 0.55 | 1.00 | 0.98 | 0.35 | 0.01 |
| Hungary | 0.44 | 0.39 | 0.11 | 0.68 | 0.56 | 0.99 | 0.97 | 0.33 | 0.01 |
| Romania | 0.50 | 0.41 | 0.07 | 0.84 | 0.63 | 1.00 | 0.97 | 0.25 | 0.01 |
| Portugal | 0.29 | 0.50 | 0.20 | 0.44 | 0.41 | 0.99 | 0.99 | 0.18 | 0.00 |
| Turkey | 0.75 | 0.43 | 0.07 | 0.86 | 0.60 | 0.95 | 1.00 | 0.09 | 0.03 |
| Russia | 0.43 | 0.46 | 0.07 | 0.93 | 0.52 | 0.95 | 0.94 | 0.34 | 0.10 |
| Brazil | 0.40 | 0.32 | 0.06 | 0.72 | 0.07 | 0.94 | 0.98 | 0.15 | 0.11 |
| Montenegro | 0.47 | 0.37 | 0.04 | 0.74 | 0.26 | 1.00 | 0.97 | 0.20 | 0.00 |
| Belarus | 0.44 | 0.35 | 0.03 | 0.90 | 0.77 | 0.93 | 0.70 | 0.00 | 0.00 |
| Egypt | 0.59 | 0.15 | 0.00 | 0.61 | 0.00 | 0.92 | 0.97 | 0.21 | 0.00 |
| Argentina | 0.59 | 0.20 | 0.12 | 0.79 | 0.35 | 0.77 | 0.74 | 0.22 | 0.01 |
| Greece | 0.00 | 0.29 | 0.19 | 0.23 | 0.25 | 1.00 | 0.99 | 0.12 | 0.00 |
| Venezuela | 0.28 | 0.00 | 0.12 | 0.79 | 0.00 | 0.00 | 0.00 | 0.32 | 0.00 |

**Table 5. *W* matrix calculated for the sample of 30 countries**

| PC1* | PC2 | PC3 | PC4 | PC5 | PC6 |
|---|---|---|---|---|---|
| 0.0037331 | 0.1495590 | 0.0802628 | 0.1350172 | 0.0779327 | 0.0581565 |
| 0.1952260 | 0.0439502 | 0.0002582 | 0.0511073 | 0.0582012 | 0.0306353 |
| 0.1771148 | 0.0104838 | 0.1769460 | 0.0015549 | 0.1137436 | 0.0310753 |
| 0.0364766 | 0.2201364 | 0.0834812 | 0.0725665 | 0.1303886 | 0.1465127 |
| 0.0986781 | 0.1619003 | 0.0091527 | 0.1438917 | 0.0726696 | 0.2638306 |
| 0.0752998 | 0.0564140 | 0.1891941 | 0.1034713 | 0.0916144 | 0.0351023 |
| 0.0785673 | 0.0574330 | 0.1900024 | 0.0780202 | 0.1206648 | 0.0985294 |
| 0.0535689 | 0.1192260 | 0.1548306 | 0.0770953 | 0.1318564 | 0.1213022 |
| 0.0154986 | 0.0669336 | 0.0888243 | 0.3118414 | 0.0039790 | 0.1098994 |
| 0.2658367 | 0.1139637 | 0.0270476 | 0.0254341 | 0.1989497 | 0.1049565 |

PC – principal component.

**Table 6. The rating vector Λ for the sample of 30 countries**

| Λ (transposed) | 0.49 | 0.16 | 0.12 | 0.09 | 0.07 | 0.03 |
|---|---|---|---|---|---|---|

**Table 7. Cluster centers for the sample of 30 countries**

| № of cluster | Rows correspond to vectors of clusters | | | | | |
|---|---|---|---|---|---|---|
| **1** | 0.259918 | 0.36576 | 0.476832 | 0.324568 | 0.368703 | 0.323598 |
| **2** | 0.068437 | 0.255082 | 0.159704 | 0.120411 | 0.180846 | 0.174925 |
| **3** | 0.541449 | 0.721616 | 0.693901 | 0.83454 | 0.625795 | 0.67663 |
| **4** | 0.468673 | 0.440418 | 0.518884 | 0.366315 | 0.507716 | 0.40509 |
| **5** | 0.455239 | 0.57652 | 0.556856 | 0.463086 | 0.546149 | 0.534525 |
| **6** | 0.852695 | 0.720394 | 0.76527 | 0.557884 | 0.816989 | 0.721204 |
| **7** | 0.70372 | 0.563184 | 0.610649 | 0.457141 | 0.661138 | 0.542587 |

**Table 8. The RELARM test model results comparison with the real country ratings**

| Country | Rating category | S&P | Moodys | Fitch |
|---|---|---|---|---|
| Switzerland | AAA | + | + | + |
| Norway | AAA | + | + | + |
| Germany | AAA | + | + | + |
| United States | AA | + | - | - |
| Austria | AA | + | + | + |
| Finland | AA | + | + | + |
| United Kingdom | AA | + | + | + |
| France | AA | + | + | + |
| Belgium | AA | + | + | + |
| Korea | AA | + | + | + |
| Czech Republic | AA | + | - | - |
| Japan | AA | - | - | - |
| China | A | - | - | + |
| Estonia | BBB | - | - | - |
| Saudi Arabia | BBB | - | - | - |
| Mexico | BBB | + | - | + |
| Kazakhstan | BBB | + | + | + |
| Bulgaria | BBB | - | + | + |
| Hungary | BBB | - | - | + |
| Romania | BBB | + | + | + |
| Turkey | BBB | - | + | + |
| Russia | BBB | - | - | + |
| Belarus | BBB | - | - | - |
| Portugal | BB | + | + | + |
| Brazil | BB | + | + | + |
| Montenegro | B | + | + | *not rated* |
| Egypt | B | + | + | + |
| Argentina | B | + | + | + |
| Greece | B | + | - | - |
| Venezuela | CCC | + | + | + |